\documentclass[prl,aps,twocolumn,showpacs]{revtex4}
\usepackage{epsfig}
\usepackage{bm}

\begin{document}

\title{Galactic turbulence and paleoclimate variability}

\author{\small  A. Bershadskii}
\affiliation{\small {ICAR, P.O.B. 31155, Jerusalem 91000, Israel}}

\begin{abstract}
The wavelet regression detrended fluctuations
of the reconstructed temperature for the past three ice ages: approximately 340000 years
(Antarctic ice cores isotopic data), exhibit clear evidences of
the galactic turbulence modulation up to 2500 years time-scales. The observed
strictly Kolmogorov turbulence features indicates the Kolmogorov nature of galactic turbulence,
and provide explanation to random-like fluctuations of the global temperature on the millennial time scales.

\end{abstract}


\maketitle

\section{introduction}

 The Earth climate is an open system. Not only the Sun is the main source of energy for the
climate dynamics but also galactic cosmic rays (GCR) can effect the climate on global scales
(see, for instance, \cite{sf}-\cite{du} and references therein). On the other hand, recent paleoclimate
reconstructions provide indications of {\it nonlinear} properties of Earth climate at the late Pleistocene \cite{rh},\cite{sal},\cite{berg1} (the period from 0.8 Myr to present). Reconstructed air temperature
are known to be strongly fluctuating on millennial time scales (see, for instance, figures 1 and 2).
While the nature of the trend is widely discussed (in relation to the glaciation cycles) the nature
of these strong fluctuations is still quite obscure. The problem has also a technical
aspect: detrending is a difficult task for such strong fluctuations.
In order to solve this problem a wavelet regression detrending method was used in present
investigation.

\begin{figure} \vspace{-1cm}\centering
\epsfig{width=.45\textwidth,file=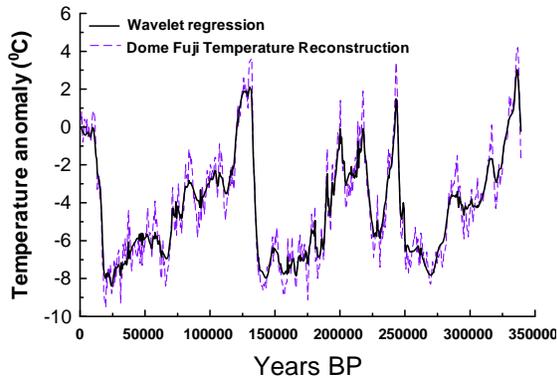} \vspace{-5cm}
\caption{The reconstructed air temperature data (dashed line) for the period 0-340 kyr.
The data were taken from Ref. \cite{Fuji} (see also Ref. \cite{ka}). The solid curve
(trend) corresponds to a wavelet (symmlet) regression of the data. }
\end{figure}
\begin{figure} \vspace{-0.5cm}\centering
\epsfig{width=.45\textwidth,file=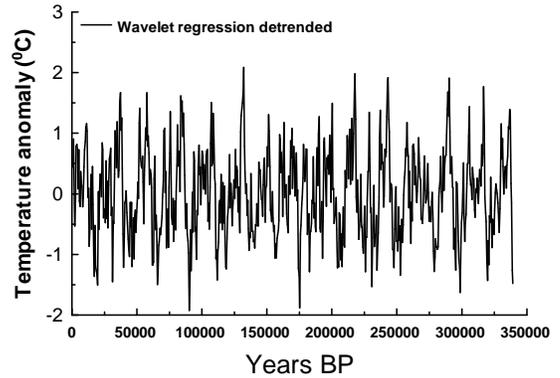} \vspace{-5cm}
\caption{The wavelet regression detrended fluctuations from the data shown in Fig. 1.}
\end{figure}

\section{Wavelet regression}

 Figure 1 shows reconstructed air temperature data (dashed line) for the period 0-340 kyr
as presented at the site \cite{Fuji} (Antarctic ice cores isotopic data, see also Ref. \cite{ka}).
The solid curve (trend) in the figure corresponds to a wavelet (symmlet) regression of the data (cf Ref. \cite{o}). Figure 2 shows corresponding detrended fluctuations, which produce
a statistically stationary set of data. Most of the regression methods are linear in responses.
At the nonlinear nonparametric wavelet regression one chooses a relatively small number of wavelet
coefficients to represent the underlying regression function. A threshold method is used to keep or
kill the wavelet coefficients. In this case, in particular, the Universal (VisuShrink) thresholding
rule with a soft thresholding function was used. At the wavelet
regression the demands to smoothness of the function being estimated are relaxed considerably in comparison
to the traditional methods.

\section{Statistical analysis}

The wavelet regression detrended fluctuations shown in Fig. 2 represent a
statistically stationary set of data and we can use such standard statistical tools as correlation function
$C(\tau)$ and  structure functions $S_p(\tau)$ (of different orders $p$) for it analysis.
The correlation function defect $1- C(\tau )$ is proportional to the second order structure function $S_2(\tau)$.
Therefore, we can compare results obtained by these different tools. First let us look at autocorrelation
function $C(\tau)$ in order to understand what happens on the millennial time scales. Figure 3 shows a relatively small-times part of the correlation function defect. The ln-ln scales have been used in this figure in
order to show a power law (the straight line) for the second order structure function: $S_2 (\tau) = \langle |x(t+\tau)- x(t)|^2 \rangle$ :
$$
1- C(\tau ) \propto  S_2(\tau) \propto \tau^{2/3}  \eqno{(1)}
$$
This power law: '2/3', for structure function (by virtue of the Taylor hypothesis
transforming the time scaling into the space one \cite{my},\cite{b1})
is known for fully developed turbulence as Kolmogorov's power law.

\begin{figure} \vspace{-0.5cm}\centering
\epsfig{width=.45\textwidth,file=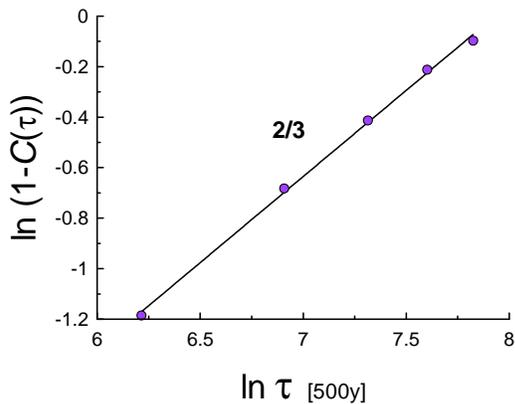} \vspace{-4.5cm}
\caption{A small-time-scales part of the autocorrelation function defect of the wavelet regression
detrended fluctuations from the data shown in Fig. 2. The straight
line indicates the Kolmogorov's '2/3' power law for the structure function in ln-ln scales.}
\end{figure}

Although, the scaling interval is short, the value of the exponent is rather intriguing.
This exponent is well known in the theory of fluid (plasma) turbulence and corresponds to so-called
Kolmogorov's cascade process. This process is very universal for turbulent fluids
and plasmas \cite{gibson}. In spite of the fact that magnetic field is presumably important for
interstellar turbulence the Kolmogorov description can be still theoretically acceptable even in this area
\cite{l},\cite{lp},\cite{clv}. Moreover, the Kolmogorov-type spectra were observed on the
scales up to kpc. In order to support the Kolmogorov turbulence as a background of the
wavelet regression detrended temperature modulation we calculated also structure functions
$S_p (\tau) = \langle |x(t+\tau)- x(t)|^p \rangle$ with different orders $p$. In the classic Kolmogorov
turbulence (at very large values of the Reynolds number \cite{my},\cite{sd})
$$
S_p \propto \tau^{\zeta_p}~~~~~~~ \zeta_p \simeq \frac{p}{3}   \eqno{(2)}
$$
(at least for $p \leq 3$). Figure 4 shows a small-time-scales part of the structure functions $S_p$ with $p=0.2,~0.5,~0.7,~1,~2,~3$
for the wavelet regression detrended fluctuations from the data shown in Fig. 2. The straight lines
are drawn in order to indicate scaling in the ln-ln scales. Figure 5 shows as circles the scaling exponent
$\zeta_p$ against $p$ for the scaling shown in Fig. 4 (the exponents were calculated using slopes of the
straight lines - best fit, in Fig. 4). The bars show the statistical errors. The straight line in Fig. 5
corresponds to the strictly Kolmogorov turbulence with $\zeta_p =p/3~$ Eq. (2). One can see good agreement with the
Kolmogorov turbulence modulation.

\begin{figure} \vspace{-0.5cm}\centering
\epsfig{width=.45\textwidth,file=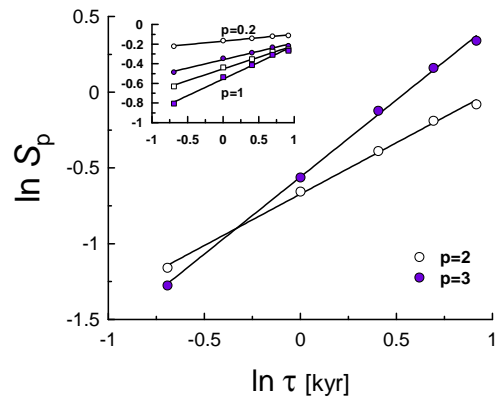} \vspace{-4.5cm}
\caption{A small-time-scales part of the structure functions $S_p$ (p=2,3) for the wavelet regression
detrended fluctuations from the data shown in Fig. 2. The insert shows the structure functions for
$p=0.2,~0.5,~0.7,~1$. The straight line indicates scaling in the ln-ln scales.}
\end{figure}
\begin{figure} \vspace{-0.5cm}\centering
\epsfig{width=.45\textwidth,file=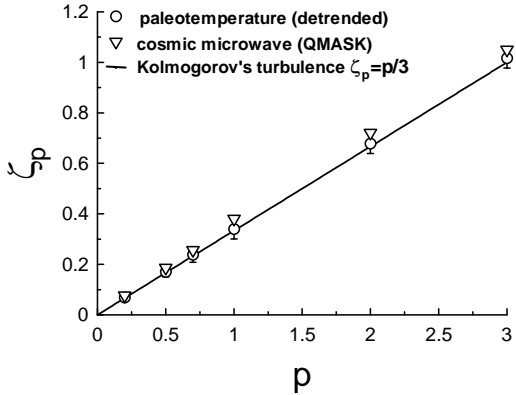} \vspace{-5cm}
\caption{The scaling exponent $\zeta_p$ against $p$ for the scaling shown in Fig. 4 (circles).
The triangles correspond to the cosmic microwave radiation (QMASK map) \cite{bs}. The straight line is
drawn in order to indicate the Kolmogorov's turbulence with $\zeta_p =p/3$ Eq. (2). }
\end{figure}

\section{Discussion}

For turbulent processes on Earth and in Heliosphere
the Kolmogorov's scaling with such large time-scales certainly cannot exist. Therefore, one should think about
a Galactic origin of Kolmogorov turbulence with such large time-scales (let us recall that diameter of the Galaxy
is approximately 100,000 light years).
This is not surprising if we recall possible role of the
galactic cosmic rays for Earth climate.  Galactic cosmic ray
intensity at the Earth's orbit is modulated by galactic turbulence \cite{b1}. On the other hand,
the galactic cosmic rays can determine the amount of cloud cover (a very significant climate factor)
on global scales through the massive aerosols formation (see, for instance, \cite{sf}-\cite{du}). Thus,
the galactic turbulence can modulate the global temperature fluctuations by the Kolmogorov
scaling properties (Figs. 3-5) on the millennial time scales. 

If one knows the characteristic velocity scale $v$ for the Taylor hypothesis one can estimate 
outer space-scale of corresponding galactic turbulence as $L \geq 2500y \times v$. 
However, it is not clear what estimate we should take for the $v$. For instance, one could try velocity of the 
solar system relative to the cosmic microwave background (CMB) rest frame: $v \sim 370$ km/sec. In this case 
one obtains $L \sim 1$pc. It should be noted that in recent paper \cite{h} it is suggested that the 
typical outer scale for spiral arms can be as small as 1pc and in interarm regions the outer scale can be larger 
than 100pc. Since the solar system and Earth are at present time within the Orion Arm this suggestion is in 
agrement with the above estimate. Although the suggestion of the Ref. \cite{h} is still under active discussion 
the paleoclimate consequences of this suggestion can be very interesting and we will discuss one of them here. 
Namely, when orbiting the Galactic center the solar system and Earth are in the interarm regions the 
reverse of the Taylor hypothesis provides us with the outer time-scale $\sim 2500 \times 100$ years. This time 
scale is larger then any known glaciation period (which are determined by the periods related to orbiting 
Earth around Sun, see for instance \cite{b}). Strong {\it turbulent} fluctuations of the cosmic rays flux on such 
large time-scales should prevent to the glaciation cycles to occur when the solar system is in the 
interarm regions. The Earth deglacitaion related to the interarm regions was suggested in Refs. 
\cite{shav1},\cite{shav2} and explained by difference in intensity of the cosmic ray flux in the spiral arms 
and in the interarm regions. It is difficult to estimate at present time which of the two mechanisms is 
more efficient. In any way they are working to the same end and both are open to discussion. \\

The same turbulence that modulates the galactic cosmic rays in the galactic disk and in its halo
\cite{clv},\cite{yl} can also modulate the cosmic microwave radiation \cite{cl}.
Naturally, the effect of the galactic modulation of the cosmic microwave radiation is different in
different regions of the sky and in different frequency bands. An additional component of
galactic microwave emission from spinning dust, for instance, has a peak in
intensity between 20 and 40 GHz \cite{dl}. The so-called QMASK map of cosmic microwave radiation
combines observations made in a vicinity of the North Celestial Pole in Ka and Q bands
(26-36 GHz and 36-46 GHz respectively) \cite{xta}. The map was generated by subdividing the sky into
square pixels of side $\Theta \simeq 0.3^{o}$ and consists of 6495 pixels.
The data are represented using three coordinates $x,y,z$ of a
unit vector ${\bf R}$ in the direction of the pixel in the map (in
equatorial coordinates). Since some pixels are much noisier than
others and there are noise correlations between pixels, Wiener
filtered maps are more useful than the raw ones. Wiener filtering
suppresses the noisiest modes in a map and shows the signal that
is statistically significant. The space increments of the cosmic microwave temperature $T$
can be defined as
$$
\Delta T_r = (T({\bf R}+{\bf r}) - T({\bf R}))   \eqno{(3)}
$$
where ${\bf r}$ is dimensionless vector connecting two pixels of the map
separated by a distance $r$, and the structure functions of order
$p$ as $\langle|\Delta T_r|^p \rangle$ where $\langle.\rangle$
means a statistical average over the map. The scaling
$$
\langle|\Delta T_r|^p \rangle \sim r^{\zeta_p}  \eqno{(4)}
$$
was observed for these increments in Ref. \cite{bs} for the QMASK map.
Corresponding values of $\zeta_p$ are shown in Fig. 5 as triangles.

\section{Acknowledgements}

The data were provided by World Data Center for Paleoclimatology, Boulder and
NOAA Paleoclimatology Program. I thank A. Lazarian for illuminating discussion 
and I acknowledge that a software provided by K. Yoshioka was used at the computations.

\end{document}